\documentclass[twocolumn]{aastex631}

\usepackage{comment,amsmath}

\shorttitle{Is the compact object associated with HESS J1731-347 a strange quark star?}
\usepackage{amsmath,amsfonts,amssymb,graphicx,color,times,bbm,psfrag,dsfont,slashed,bigints,bm}

\begin{document}
\title{Is the compact object associated with HESS J1731-347 a strange quark star?\\A possible astrophysical scenario for its formation}

\author[0000-0002-8257-3819]{Francesco Di Clemente}
\affiliation{Dipartimento di Fisica e Scienze della Terra dell’Universit\`a di Ferrara, Via Saragat 1,
I-44122 Ferrara, Italy}
\affiliation{INFN Sezione di Ferrara,
Via Saragat 1, I-44122 Ferrara, Italy}

\author[0000-0003-1302-8566]{Alessandro Drago}
\affiliation{Dipartimento di Fisica e Scienze della Terra dell’Universit\`a di Ferrara, Via Saragat 1,
I-44122 Ferrara, Italy}
\affiliation{INFN Sezione di Ferrara,
Via Saragat 1, I-44122 Ferrara, Italy}

\author[0000-0003-3250-1398]{Giuseppe Pagliara}
\affiliation{Dipartimento di Fisica e Scienze della Terra dell’Universit\`a di Ferrara, Via Saragat 1,
I-44122 Ferrara, Italy}
\affiliation{INFN Sezione di Ferrara,
Via Saragat 1, I-44122 Ferrara, Italy}

\begin{abstract}
The analysis of the central compact object within the supernova remnant HESS J1731-347 suggests that it has a small radius and, even more interestingly, a mass of the order or smaller than one solar mass.
This raises the question of which astrophysical process could lead to such a small mass, since the analysis of various types of SN explosions indicate that is it not possible to produce a neutron star with a mass smaller than about $1.17 M_\odot$.
Here we show that masses of the order or smaller than one solar mass can be obtained in the case of strange quark stars and that it is possible to build a coherent model explaining not only the mass and the radius of that object, but also its slow cooling suggested in various analyses. 
We also show that an astrophysical path exists which leads to the formation of such an object, and we discuss the role played in that scenario by strangelets assumed to constitute the dark matter.

\end{abstract}

\section{Introduction}
The analysis of the central compact object within the supernova (SN) remnant HESS J1731-347 suggests that it has a small radius and, even more interestingly, a mass of the order or smaller than one solar mass \citep{doroshenko2022}. This raises the question of which astrophysical process could lead to such a small mass, since the analysis of various types of SN explosions indicate that is it not possible to produce a neutron star (NS) with a mass smaller than about $1.17 M_\odot$ \citep{Suwa:2018uni}. First, we will show that masses of the order or smaller than one solar mass can be obtained in the case of strange quark stars (QSs) due to their large binding energy, and that it is possible to build a coherent model explaining not only the mass and the radius of that object, but also its slow cooling suggested in various analyses \citep{Klochkov:2014ola,Potekhin:2020ttj}. Moreover, we will show that QSs can fulfill all the limits on masses and radii of the other astrophysical objects discussed in \citet{doroshenko2022} and can also explain the possible existence of objects having a mass of the order or larger than $2.5 M_\odot$, as suggested by the analysis of GW190814 \citep{LIGOScientific:2020zkf,Bombaci:2020vgw}.
Finally, we will discuss the astrophysical processes that can lead to the production of such an object. We will consider the mechanism  based on progenitors having a mass in the range $(8-10) M_\odot$ \citep{Suwa:2018uni} and we will show that a crucial role is played by the hypothesis that dark matter (DM) is made of strange quark matter \citep{witten84}, since the presence of strange quark matter, captured by the progenitor dense core during the last stages before collapse and present inside the proto-neutron star can trigger the conversion of that object into a QS \citep{Madsen:1988zgf}.

\section{Binding energy}
In order to understand why QSs can be produced with a mass of the order or smaller than one solar mass, it is useful to recall that the total binding energy (BE) of compact stars is the sum of the gravitational BE and the nuclear BE, the latter being related to the microphysics of the interactions as discussed in the seminal paper \citet{Bombaci:2000cv}.
While the first is positive (binding) both for NSs and for QSs, the second is large and negative for NSs (anti-binding) and either small and negative or positive for QSs \citep{Drago:2020gqn}. In this way the total BE of QSs can be significantly larger than that of NSs. This important point was used in our recent paper, \citet{DiClemente:2022ktz}, where we suggested a mechanism to produce compact objects having a mass smaller than one solar mass via accretion induced collapse of white dwarfs containing strange quark matter at their center. Clearly, this mechanism for producing QSs does not apply to the case of HESS J1731-347, since the central object is embedded in a dust shell of $\sim 2 M_\odot$ expelled at the moment of the SN explosion \citep{Doroshenko2016}. On the other hand, the large difference between baryonic ($M_b$) and gravitational mass ($M_g$) in the case of QSs is a general property that can also be applied to other scenarios. In the Table we show examples of total BEs of QSs and of NSs. In particular, in the NS case, we use the empirical relation suggested in \citet{Lattimer:2000nx}: $\mathrm{BE}/M_\odot=0.084 (M_g/M_\odot)^2$ which
correctly describes the results for a large class of hadronic equations of state.
\begin{table}[t]
\centering
\begin{tabular}{ccccc}
\hline
\rule[-3mm]{0mm}{7mm}
$M_b $& $M_{g}^{NS} $  & $M_{g,A}^{QS} $ & $M_{g,B}^{QS} $ & $M_{g,C}^{QS} $  \\ \hline
$1.28$ &$1.17$& $0.99$ & $1.00$ & $0.95-1.05$  \\
$1.32$  & $1.20$ & $1.01$  & $1.03$ & $0.98-1.08$ \\ \hline
\\
\end{tabular}
\caption{Minimum allowed mass (in units of $M_{\odot}$) for NSs and for QSs in three models. \textit{A} refers to the EoS in \cite{Bombaci:2020vgw} (solid red line in the Figure). \textit{B} refers to a EoS derived in \cite{Ferrer:2015vca} (solid blue line). \textit{C} refers to the most probable EoS having a constant speed of sound and is obtained from the bayesian analysis in \cite{Traversi:2021fad} which does not include the most recent data on massive stars (solid black line). In the latter case a range of values is indicated, since the BE is not fixed by the bayesian analysis. The chosen values correspond to an energy per baryon of strange quark matter at zero pressure of $(E/A)_{p=0}=(765-850) \mathrm{MeV}$, in agreement with the discussion in \cite{Weber:2004kj}.}
\label{tab:minimummass}
\end{table}
In \cite{Suwa:2018uni} it has been shown 
that the minimum baryonic mass of the core of the progenitor of a SN is in the range $(1.32-1.28 )M_\odot$. When using the previous parameterisation of the total BE of NSs, this corresponds to a $M_g$ in the range $(1.20-1.17) M_\odot$. Instead, as shown in the Table, the $M_g$ of a QS having a baryonic mass in that range can be of about one solar mass or slightly smaller, which provides an explanation of the small mass proposed in \citet{doroshenko2022}.

\section{Masses, radii and cooling}

\subsection{Mass-radius relation}
\begin{figure}[t]
\centering

\begin{minipage}{0.5\textwidth}
\includegraphics[width=0.9\textwidth]{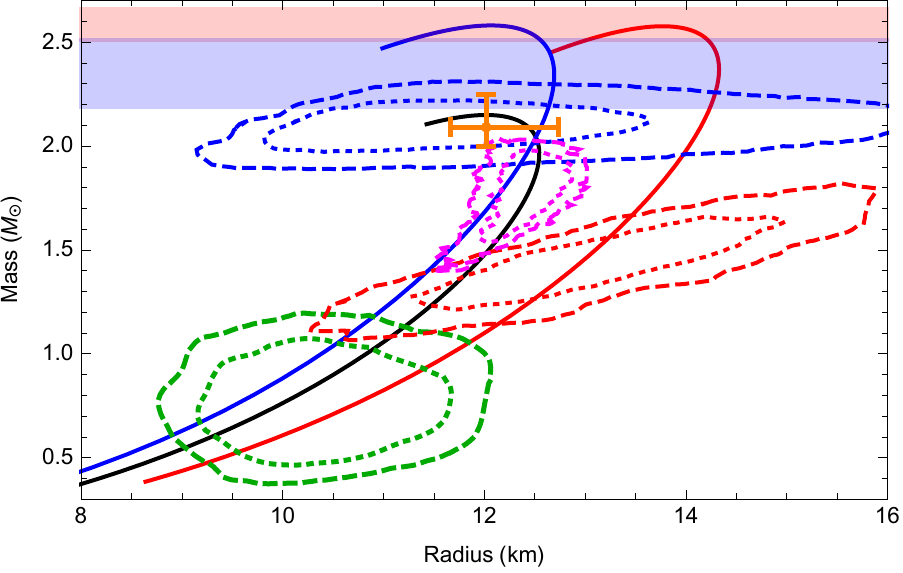}
\end{minipage}
\caption{Mass-radius relation of QSs from \cite{Bombaci:2020vgw} (solid red), \cite{Ferrer:2015vca} (solid blue) and \cite{Traversi:2021fad} (solid black) with observational constraints at 68\% of confidence level (dotted) and at 90\% (dashed). Blue: analysis of PSR J0740+6620 from NICER and XMM-Newton data from \cite{Miller:2021qha}. Magenta: analysis of 4U 1702-429 from \cite{Nattila:2017wtj}. Red: analysis of PSR J0030+0451 from \cite{Riley:2019yda}. Green: latest analysis of HESS J1731-347 from \cite{doroshenko2022}. Orange error bars: analysis of 3XMM J185246.6+003317 from \cite{deLima:2022hje}.} 
\label{fig:mrconstraints}
\end{figure}

In Fig.\ref{fig:mrconstraints} we display a variety of astrophysical limits. We have added to those discussed in \cite{doroshenko2022} the limits on the mass stemming from the analyses of the light component of GW190814 and of PSR J0952-0607, and the limits on mass and radius of 3XMM J185246.6+003317 \cite{deLima:2022hje}. We also display three examples of mass-radius relation of QSs taken from \cite{Bombaci:2020vgw,Traversi:2021fad,Ferrer:2015vca}: these can satisfy all the constraints, including those for large masses.
The possibility of reaching large masses (and radii) with QSs has been suggested over many years \cite{Alford:2006vz,Kurkela:2009gj}, but the upper mass limit has to be defined by future and more conclusive observations.

We have shown that QSs can explain the existence of compact stars having very small or very large masses. On the other hand, it is unlikely that all compact stars are QSs: it is well known that magnetar oscillations pose challenges to QSs \citep{Watts:2006hk}. Also, the analysis of the energy released by the SN1987a indicates a binding energy perfectly compatible with that of a NS \citep{Pagliaroli:2008ur}. Instead, in order to satisfy the limit on the mass of HESS J1731-347, we have explicitly used the significantly larger value of the binding energy of a QS. In the last decade we have developed a scheme, named two-families scenario, in which NSs and QSs co-exist \citep{Berezhiani:2002ks,Drago:2014oja} and we have discussed in several papers the pathways leading to the formation of a QS \citep{Wiktorowicz:2017swq,DePietri:2019khb,DiClemente:2022ktz}. The large range of masses of QSs does not therefore rule out the possible existence of NSs. 

\subsection{Thermal evolution}
An important feature of the central object inside HESS J1731-347 is that it cools down very slowly, with a surface temperature $\mathrm{T_s}\sim 153^{+4}_{-2}\, \mathrm{eV}$  and an age in the range $(2-6)\, \mathrm{kyr}$ \cite{Potekhin:2020ttj}. Moreover, \citet{Doroshenko2016} estimate that this object has an age of $\sim4$ kyr, based on modeling of the companion star. This suggests a thermal evolution similar to that of a standard NS. On the other hand, we are claiming that the object is a QS and, before the discovery of color superconductivity of quark matter, it was believed that the temperature of QSs would drop much more rapidly than that of NSs. Actually, the formation of gaps suppresses the rapid cooling mechanisms in quark matter, which is similar to what happens in hadronic matter.Therefore, the cooling curves of QSs and of slowly cooling NSs can be almost indistinguishable \cite{Schaab:1997hx,Weber:2004kj}.
In \cite{DiClemente:2022ktz} we have suggested that SAX J1808.4-3658 is also a QS, since it has been indicated that its mass could be smaller than one solar mass \cite{DiSalvo:2018mua}. On the other hand, the thermal emission of that object indicates that some form of enhanced cooling takes place in the star \cite{Heinke:2008vj}. There are possible explanations of the very different behaviour of these two objects. First, a carbon heat blanket can be present in the case of HESS J1731-347, making the surface hotter \cite{Klochkov:2014ola}. Also, it has been shown that a transition between slow and rapid cooling for stars containing quark matter can take place at a critical temperature (related to the formation of specific phases in quark matter) and the transition can be extremely rapid \cite{Sedrakian:2013pva}. It is therefore possible that the slow cooling object in HESS J1731-347 is slightly younger than the rapid cooling SAX J1808.4-3658. Finally, SAX J1808.4-3658 is accreting mass and it produces powerful outbursts (while the object at the center of HESS J1731-347 is completely quiet) during which it displays a highly variable luminosity, which could be explained if the temperature during the outbursts increases enough to again exceed the critical temperature \cite{Sedrakian:2015qxa}.

\section{Astrophysical path}
We need to identify a possible astrophysical path leading to the production of such a subsolar-mass object. In \citet{Suwa:2018uni} evolved stars in the mass range $(8-10) M_\odot$, having a carbon-oxygen core, have been suggested as possible progenitors of low mass NSs. We will follow that same idea and we will clarify the conditions under which a QS is produced instead of a NS.

In previous articles we have proposed a few astrophysical processes leading to the formation of a QS \citep{Drago:2015dea}. The general idea is that strange quark matter (if not already present) can form if a large density of strange hadrons is present in the object \citep{Berezhiani:2002ks,Bombaci:2004mt,Bombaci:2016xuj}. QSs can form, for instance, through mass accretion onto NSs, since the central density increases and hyperons can be produced \citep{Wiktorowicz:2017swq}; in the merger of two compact stars \citep{Drago:2015qwa,DePietri:2019khb}, due to both the increase of density and temperature; in core collapse SNe if large enough densities and temperatures are reached \citep{Drago:2015dea}. In all those cases the density of hyperons reaches a critical value above which thermal or quantum nucleation of bubbles of strange quark matter can take place.

The astrophysical processes discussed in \citet{Suwa:2018uni} are associated with explosion mechanisms in which the densities and temperatures reached are not too large \citep{Leung:2019phz}. At densities smaller than about twice nuclear matter saturation density and entropies in the central region $S/k_\mathrm{B} \lesssim 2$, the hyperon fraction is negligible and quark nucleation will not start \citep{Bombaci:2011mx}. Therefore, we do not think that a clear mechanism exists, leading to the formation of a QS for those objects {\it{unless}} some amount of quark matter is already present at the center of the progenitor: in that case, the germ of quark matter already present can trigger the deconfinement of the entire NS. 

Nuggets of strange quark matter can be produced through the ejecta of QS-QS mergers \citep{Bauswein:2008gx,Bauswein:2009im} or at the time of cosmological hadronization, in which case they can constitute the DM \citep{witten84}.
The rather limited number of strangelets produced in QS-QS mergers makes them difficult to detect, and they do not play a crucial role in stellar evolution
\citep{Bucciantini:2019ivq}. Instead, if strangelets constitute DM their flux is large and they can accumulate inside a star and influence its evolution \citep{Madsen:1988zgf}.

This possibility has been discussed many times in the past, starting from the seminal papers of \citet{witten84} and \citet{Madsen:1988zgf} to more recent review papers such as \citet{Burdin:2014xma,Jacobs:2014yca}. Also, the search for nuggets of strange quark matter is still very active \citep{Bacholle:2020emk,POEMMA:2020ykm,JEM-EUSOColaboration:2014pci} and the observational limits on their existence are continuously updated. The more recent limits indicate that, if strangelets constitute most of the DM, the peak of the distribution of their masses should be in the range $10^{10}\mathrm{g}\lesssim m\lesssim 10^{18}\mathrm{g}$ \citep{SinghSidhu:2019tbr}. While other mass windows exist in general for "macroscopic" DM, they are closed in the specific case of strangelets because they would affect stellar evolution in ways which are ruled out by observations \citep{Madsen:1988zgf}. It is interesting to note that the allowed mass range is compatible with that suggested in the seminal paper of \citet{witten84}.
 
In the following we will try to estimate the capture rate of strangelets by stellar objects and in particular by main sequence stars having developed a degenerate core. Let us start by using the simple scheme outlined in \citet{Madsen:1988zgf} since it constitutes an important reference point despite being based on a few simplifying assumptions.  If we assume for strangelets the same density as 
 that of DM in the galactic halo $\rho_\mathrm{DM}\sim 10^{-24}\,\mathrm{g/cm^3}$ and a velocity of $250\, \mathrm{km/s}$, the rate of impact of strangelets having baryonic number A onto stellar objects of mass $M$ and radius $R$ reads \citep{Madsen:1988zgf}:
\begin{equation}
F\sim (1.39 \times 10^{30} \mathrm{s}^{-1}) A^{-1}(M/M_\odot)(R/R_\odot)\, , \label{madsen}
\end{equation}
where we have assumed that all strangelets have the same baryon number $A$.
We will concentrate on the capture rate onto the dense core of evolved stars and, in particular, of stars with a zero-age-main-sequence mass of $\sim (8–10) M_\odot$, which can be the progenitors of low mass NSs \citep{Suwa:2018uni}. Their core (and, more generally, the core of the progenitor of a NS produced in a core-collapse) is similar to a white dwarf with a mass $M\sim (1.3-1.4) M_\odot$ and a radius $R\sim 10^{-2} R_\odot$. By using Eq.(\ref{madsen}), the capture rate on the core of those stars reads:
\begin{equation}
F_{\mathrm{c}}\sim 6 \times 10^{35} \mathrm{yr}^{-1} A^{-1}\sim 10^{12} \mathrm{yr}^{-1} m_s^{-1}\, , \label{cr}
\end{equation}
where $m_s$ is the mass of the strangelet in grams. This corresponds to about one capture per year for strangelets of $\sim 10^{12}$g, which is the typical mass discussed in the following.

In principle, a QS can be produced also via the capture of a strangelet by an already formed NS. In particular, this can happen during the first month, before a crust is formed \citep{Madsen:1988zgf}. Notice anyway that, if this mechanism is possible, then all compact stars are QSs, what is in disagreement with observations. In \citet{Bucciantini:2019ivq} we have shown that the strangelets produced by the merger of two QSs are too few to be captured, and in \citet{diclemente2024strange} we have shown that the distribution of strangelets constituting DM also can avoid that conversion. 

We now want to evaluate the capture rate in a more precise way, by assuming that the strangelets have different sizes, with a mass distribution peaked in the range indicated above, which corresponds to the range $10^{34}\lesssim \mathrm{A} \lesssim 10^{42}$. The capture rate depends on three factors: 

\begin{enumerate}
    \item the position of the object in the galaxy;
    \item the time during which the dense core exists before the collapse;
    \item the distribution of the sizes of the strangelets.
\end{enumerate}
Concerning the first point, the density of DM in the galactic bulge can easily be 1-2 orders of magnitude larger than in the area where the Sun is located \citep{Nesti:2013uwa,Salucci:2018hqu}, and this can be a relevant factor, as we already suggested in the case of accretion-induced collapse of strange dwarfs \citep{DiClemente:2022ktz}. Concerning the time interval between the formation of a dense core inside the progenitor and the moment of its collapse, it depends on the mass of the progenitor. Following \citet{Madsen:1988zgf}, we assume that the core of the progenitor is dense enough to stop and capture the strangelets if silicon has been produced, since in that case the central density of the core reaches about $10^8\, \mathrm{g/cm^3}$. The typical time scale $\tau(M)$ is of the order of a year \citep{arnett,kippenhahn} and depends on the mass of the star, the more massive ones evolving faster. 

\begin{figure}[t]
\centering
\begin{minipage}{0.5\textwidth}
\includegraphics[width=0.9\textwidth]{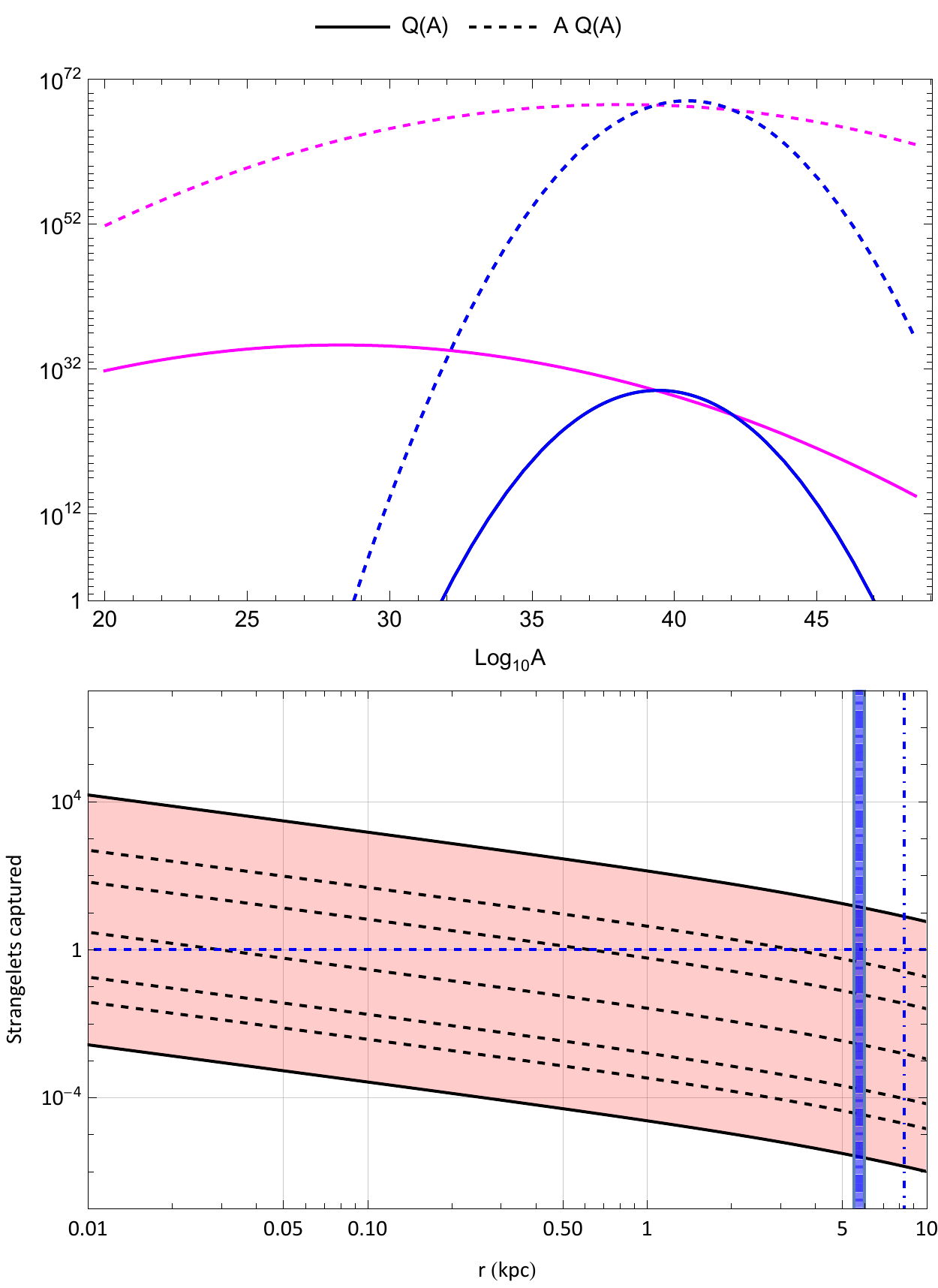}
\end{minipage}
\caption{Top panel: distributions of the number density of fragments $Q(A)$ and of the baryonic mass $A \,Q(A)$, as functions of $\mathrm{Log}_{10}A$. The distributions in magenta (blue) correspond to a total number of strangelets in the Milky Way of $\sim 10^{36}$ ($\sim 2\times 10^{29}$).  Bottom panel: strangelets that can be captured from the core of stars having 10 $M_\odot$. The limiting solid curves correspond to the two distributions in the top panel while the dashed black lines are intermediate value of the total number of strangelets (from the bottom $ 2.6 \times10^{30}$, $1.2 \times10^{31}$, $2.0 \times10^{32}$, $4.6 \times10^{33}$, $3.4 \times10^{34}$). The vertical dashed line represents the Earth position, while the horizontal line represents the threshold beyond which at least one strangelet is captured by the silicon core of a 10 $M_\odot$ star. The vertical band represents the position of HESS J1731-347.}
\label{fig:capturedstrangelets}
\end{figure}

In \citet{diclemente2024strange} an attempt has been made to compute the distribution of the masses of the strangelets produced in the early Universe, under the assumption that they constitute the DM. A rather simple scheme for the formation of cosmological strangelets have been discussed, based on the following ideas: 1) assume that the clusters of strange quark matter form through coalescence of smaller pre-clusters, and follow the dynamics of these objects to estimate their initial size \citep{witten84}, which turns out to be of the order of centimeters; 2) take into account the partial evaporation of the strangelets, which happens when the temperature drops below $\sim$ 100 MeV; below that temperature the neutrinos can propagate in an optically thin region around the strangelets and keep the strangelets hot, as discussed in \citet{Alcock:1985vc,Madsen:1986jg}. This procedure is also similar to that followed in \citet{Bucciantini:2019ivq}. Notice that the evaporation ends when the temperature reaches about 10 MeV, well above the temperature at which nucleosynthesis takes place.

The strangelet distributions must respect the limits on the masses discussed above and also those set by the observed femtolensing \citep{Burdin:2014xma} which indicates that the contribution to the DM fraction for objects in the range of mass from $5 \cdot 10^{17}\, \mathrm{ g}$ to $2 \cdot 10^{20}\, \mathrm{ g}$  has to be less than 8\%\citep{Barnacka:2012}. Moreover, a newly formed NS could convert to a QS by capturing a strangelet, before forming a crust \citep{Madsen:1988zgf} and if that happens NSs could not exist. To avoid this problem, we impose that, at least at a distance larger than $1 \,\mathrm{ kpc}$ from the galactic center, the probability that a newly-formed NS captures a strangelet is $< 1$.
As discussed in \citet{diclemente2024strange}, it is possible to fix the few parameters in this scheme so as to have the peak of the mass distribution within the allowed range indicated above. Moreover, we can reproduce the correct ratio between luminous and dark matter, that in this scenario corresponds to the ratio between the evaporated/not evaporated part of the original quark clusters. In the upper panel of Fig.\ref{fig:capturedstrangelets} we show two examples of the mass distributions of the strangelets obtained by using the scheme described above. 

We can now estimate the number of strangelets collected by the dense core of the progenitor in a more precise way than by using Eqs.(\ref{madsen}-\ref{cr}). We use a modified version of Eq.~3 of \citet{Jacobs:2014yca}, taking into account that not all the strangelets interacting gravitationally with the star are stopped, but only those that impact on the high-density central region of radius $R_c$. The capture rate reads therefore: 
\begin{align}
F_c=&(2.7 \times 10^{29} \mathrm{s}^{-1})\,\frac {\rho_\mathrm{NFW}(r)}{\rho_\mathrm{DM}}\, \frac {N_\mathrm{S}} {M_\mathrm{V}} \frac{M}{M_\odot}\, v_{250}\, \left(\frac{R_c}{R_\odot}\right)^2 \nonumber \\ &\times \left(1+ 6.2 \left(\frac{R_\odot}{R_t}\right) \,  v_{250}^{-2}\, \left(\frac{M_t}{M_\odot}\right) \right)\label{eq:jacobs}
\end{align}
where $\rho_\mathrm{NFW}(r)$ is the Navarro-Frank-White profile \citep{Navarro:1995iw,Nesti:2013uwa}, $N_\mathrm{S}$ is the total number of strangelets in the galaxy, $M_\mathrm{V}$ is the virial mass of the Milky Way in units of the proton mass, $v_{250}$ is the velocity of the DM in units of 250 km/s, $M$ is the mass of the star, R$_c$ is the radius of the star core (which is the part of the star capable of stopping strangelets \citep{Madsen:1988zgf}).
Notice that Eq.~\ref{eq:jacobs} reduces to Eq. \ref{madsen} if all the strangelets have the same mass, if the geometrical cross-section is negligible respect to the gravitational one and if the gravitational radius $R_t$ equals the capture radius $R_c$.

Concerning HESS J1731-347, it is located in one of the Galactic spiral arms along the line of sight \citep{Landstorfer2022}, in the direction of the galactic center at a distance from the Sun of $\sim$ 2.5 kpc \citep{doroshenko2022}. The density of DM in that region can be higher than the  density at the Sun position up to a factor of $\sim 5$ \citep{Portail:2016vei}.
In the bottom panel of Fig.~\ref{fig:capturedstrangelets} we present the number of strangelets that can be captured by a 10 $M_\odot$ progenitor, assuming the NFW profile for the DM in the Milky Way \citep{Nesti:2013uwa}. The results for a $25 M_\odot$ progenitor (not shown in the figure) are almost indistinguishable from the case of 10 $M_\odot$ because the increase in the capture rate due to the larger mass is compensated by the decrease in $\tau(M)$. It is clear that a rather large range of possible distributions of strangelets exists using which only the proto-NSs closer to the galactic center will convert to QSs and not all the NSs in galaxy. 

\section{Conclusions}
We have shown that QSs can successfully explain the properties of a few compact objects and, in particular, the central object in HESS J1731-347. We have also shown that NSs can co-exist with QSs since it is possible that the conversion of NSs to QSs due to DM takes place mostly in the central region of the galaxy. This co-existence corresponds to the so-called two-families scenario \citep{Berezhiani:2002ks,Drago:2013fsa}, which also remains valid when strangelets produced by the merger of two QSs are taken into account \citep{Bucciantini:2019ivq}. 

The core of the scenario here proposed is that subsolar compact objects can be QSs, and that their production necessitates the pre-existence of a seed of strange quark matter in the progenitor. This is true both in the case of the mechanism here discussed, based on the collapse of progenitors having a mass in the range $(8-10) M_\odot$, and in the scenario discussed in \citep{DiClemente:2022ktz} based on the accretion induced collapse of strange dwarfs. We have shown that the presence of the quark seed can be justified if DM is made of strange quark matter \citep{witten84,diclemente2024strange}. In turns, this suggests that subsolar mass compact stars should be more frequent in the core of galaxies, where DM is more abundant.

Finally, it is interesting to note that the conversion to QSs can also take place for proto-neutron stars whose progenitors have a mass larger than 10 $M_\odot$. This poses the question of which signatures can be associated with the deconfinement of quarks in core collapse SNe. Since the energy released in the conversion is extremely large \citep{Drago:2004vu,Bombaci:2020vgw}, it is natural to associate the formation of QSs with very energetic events, such as the superluminous SNe \citep{Moriya:2018sig}. 

\bibliography{ref}
\end{document}